\documentclass[aps,pb,amsmath,amssymb,twocolumn,twoside,showpacs,superscriptaddress,longbibliography]{revtex4-2}
\usepackage[pdftex]{graphicx}
\usepackage[colorlinks=true,citecolor=blue]{hyperref}
\usepackage{graphicx,textcomp,dcolumn,hyperref,upgreek}

\usepackage{physics}

\usepackage{url}

\usepackage{lipsum}

\usepackage{dsfont}

\usepackage{blindtext}

\usepackage{MnSymbol}  

\newcommand{\z}{_{\noindent z}}

\newcommand{\bp}{z_\text{A}}

\begin{document}

\title{
Dark versus blocking states in electronic transport: \\
a Lee-Yang zero analysis of full counting statistics
}
\author{Johann~Z{\"o}llner}
\email{johann.zoellner@uni-due.de}
\affiliation{Faculty of Physics and CENIDE, University of Duisburg-Essen, 47057 Duisburg, Germany}

\author{Philipp~Stegmann}
\affiliation{Department of Chemistry, Massachusetts Institute of Technology, Cambridge, Massachusetts 02139, USA}

\author{J{\"u}rgen~K{\"o}nig}
\affiliation{Faculty of Physics and CENIDE, University of Duisburg-Essen, 47057 Duisburg, Germany}

\author{Eric~Kleinherbers}
\affiliation{Faculty of Physics and CENIDE, University of Duisburg-Essen, 47057 Duisburg, Germany}
\affiliation{Department of Physics and Astronomy, University of California, Los Angeles, California 90095, USA}

\date{\today}

\begin{abstract}
    Electronic transport through nanostructures can be suppressed by coherent population trapping, in which quantum coherence leads to a \textit{dark state} that decouples from the drain electrode.
    Finite transport, then, relies on decoherence of the dark state.
    An alternative scenario for reduced transport is weak coupling of a state, referred to as a \textit{blocking state}, to the drain.
    This raises the question of whether and how these two scenarios can be distinguished in the transport features. For the example of electron transport through a carbon nanotube we analyze the full counting statistics in terms of Lee-Yang zeros and factorial cumulants. This allows us to identify regimes in which the distinction between dark and blocking state is possible and regimes in which this is not the case.
\end{abstract}

\maketitle

\section{Introduction}{\label{sec:Introduction} 

The phenomenon of \textit{coherent population trapping} occurs in $\Lambda$-type atomic systems, when laser fields drive the atom into a particular linear combination of eigenstates, a so-called \textit{dark state}, which becomes fully transparent to the light~\cite{alzetta_experimental_1976,arimondo1976nonabsorbing,arimondo_coherent_1996,scully_1999}.
Absorption and subsequent fluorescent emission of light is, then, only possible after decoherence of the dark state.
All-electric analogs of coherent population trapping have been proposed in quantum-dot systems \cite{brandes_current_2000,Michaelis_2006,groth_counting_2006,poeltl_twoparticle_2009,payette_coherent_2009}, and an experimental verification of this effect in a carbon nanotube has been recently reported in Ref.~\cite{donarini_2019}. 
Electron transport in nanostructures is, however, often subject to various competing microscopic mechanisms that may complicate the identification of a particular effect.
A suppressed current through a carbon nanotube or a quantum dot may be due to coherent population trapping, but it can also arise in the presence of a so-called \textit{blocking state}, i.e., a state that is just weakly coupled to the drain electrode~\cite{harabula_2018}.
While for a dark state decoherence and/or virtual charge fluctuations are required to sustain a finite remaining current, electrons in a blocking state are directly coupled to the drain electrode, even if only weakly.

This raises the question of how to distinguish a dark-state from a blocking-state model based on transport measurements.
Since both scenarios imply a suppressed current, the suppression alone is not sufficient to prove coherent population trapping.
More information about the underlying dynamics and transport mechanisms is contained in the current fluctuations.
We, therefore, analyze the full counting statistics of electron transfers.
Such a strategy has been successful for experimentally accessing spin relaxation in singly-charged quantum dots~\cite{kurzmann_2019}, the interaction between two spin-crossover molecules in nanojunction~\cite{stegmann_kemp_2021}, a long-lived coherence between spin excitations in antiferromagnetic complexes attached to a carbon nanotube~\cite{besson_2023}, and spin-flip Raman and Auger processes in self-assembled quantum dots~\cite{kleinherbers_2023_unraveling}.

\begin{figure}[t]
  \begin{center}
    \includegraphics[width=0.5\textwidth]{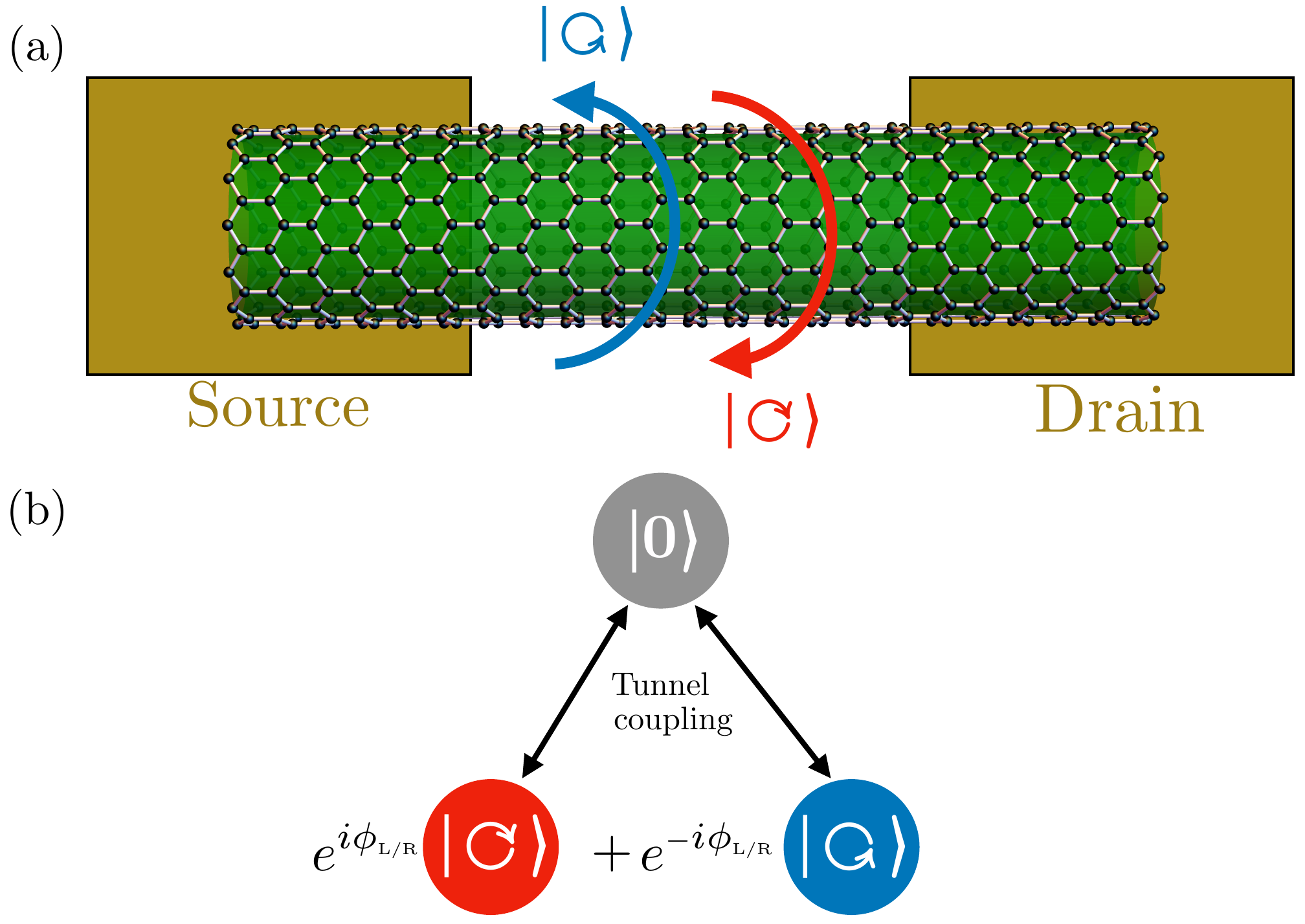}
  \end{center}
  \vspace{-20pt}
  \caption{(a) A carbon-nanotube quantum dot coupled to source and drain. Figure adapted from Ref.~\cite{donarini_2019}.
(b) Analog of an optical $\Lambda$ system, where the tunnel-coupled source ($\Gamma_\text{L}$) and drain ($\Gamma_\text{R}$) enable transitions between the empty state $\ket{0}$ and particular linear combinations of the valleys, $e^{i\phi_\text{L/R}}\ket{\lcirclearrowright}+e^{-i\phi_\text{L/R}}\ket{\rcirclearrowright}$.}
\label{fig:setup}
\end{figure}

For our study, we choose the same model system as in Ref.~\cite{donarini_2019} and use system parameters close to the experimental values. 
The experimental setup is sketched in Fig.~\ref{fig:setup}(a). 
A quantum-dot device is realized by a carbon nanotube suspended on top of two leads depicted in gold. 
The left lead is slightly rotated along the tube axis with respect to the right lead (not visible), resulting in different tunnel couplings for the quantum-dot states to source and drain. This enables an excitation of a special linear combination of eigenstates, which completely decouples from the drain electrode, thus forming a dark state.
As a consequence, once this dark state is occupied, the quantum dot is coherently trapped and electron transport is suppressed.
Only with a small probability, the dark state can change, either by decoherence or by virtual charge fluctuations with the source electrode, to another state that is coupled to the drain electrode, and electron transport is resumed.

We contrast this \textit{dark-state} model with an alternative one, the \textit{blocking-state} model, in which transport is carried through a state that is only weakly coupled to the drain electrode.
In this case, it is not the coherence but the weak coupling that is responsible for the suppression of current.
We aim at distinguishing these two models by analyzing the full counting statistics of electron transfer.
As a theoretical indicator of qualitatively different transport behavior, we employ the Lee-Yang zeros of the moment generating function in the complex plane.
The position of the Lee-Yang zeros affects the so-called factorial cumulants, that are experimentally accessible.
In particular, the sign of the factorial cumulants will be used as an indicator.

This paper is organized as follows. 
We introduce the dark-state and the blocking-state model in Sec.~\ref{sec:system} and discuss their dynamics in terms of a kinetic equation in Sec.~\ref{sec:kinetic}.
Then, in  Sec.~\ref{sec:factorial_cumulants}, we analyze the full counting statistics in terms of factorial cumulants.
To get a deeper insight, we use the Lee-Yang zeros in Sec.~\ref{sec:lee_yang} to distinguish qualitatively different topologies of their arrangement in the complex plane.
As a result, we find that for the system parameters of Ref.~\cite{donarini_2019}, the dark state and the blocking states cannot be distinguished by means of full counting statistics.
We identify, however, a regime with different system parameters in which the two models can be clearly discriminated from each other by the sign of the factorial cumulants.
Finally, we conclude our findings in Sec.~\ref{sec:conclusions}.

\section{System}\label{sec:system}

\begin{figure}[t]
  \begin{center}
    \includegraphics[width=0.5\textwidth]{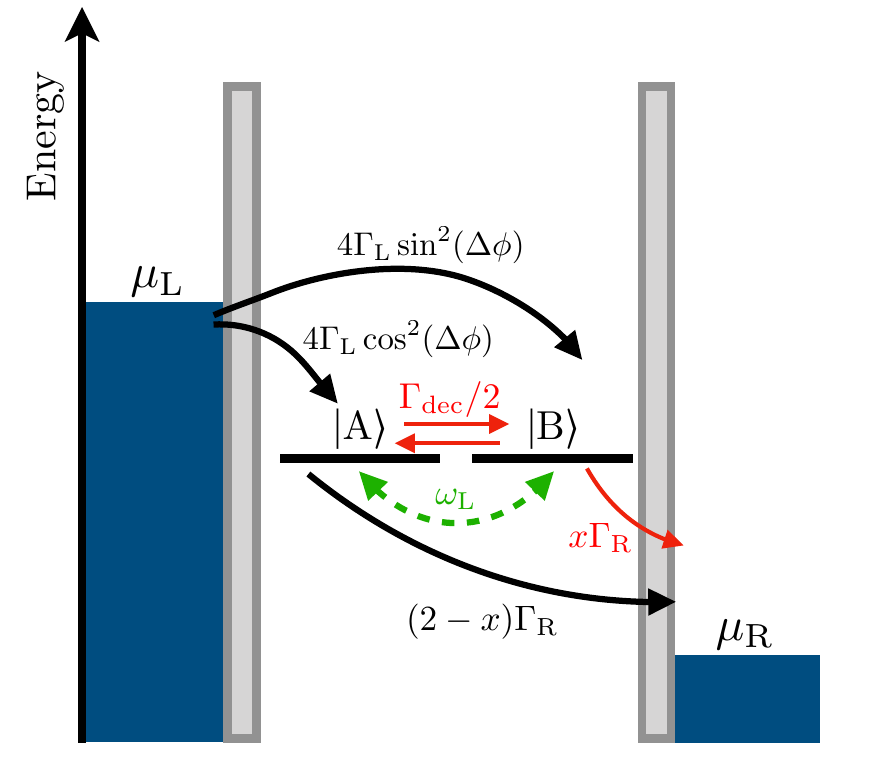}
  \end{center}
  \vspace{-20pt}
  \caption{Excitation energies for the quantum-dot system. The state $\ket{\text{B}}$ can be excited by the left lead, but does not couple to the right lead if $x=0$. In this case, it becomes a dark state. Thus, electron transport is only possible through the coupled state $\ket{\text{A}}$, via a prior transition from $\ket{\text{B}}$ to $\ket{\text{A}}$ with the decoherence rate $\Gamma_\text{dec}$ or via a coherent evolution $\omega_\text{L}$ induced by the left lead. For $x>0$, the state $\ket{\text{B}}$ becomes a blocking state, which is weakly coupled to the right lead.}
  \label{fig:energies}
\end{figure}

For the regime, we are interested in (shell 1 in Ref.~\cite{donarini_2019}), the quantum dot can be modeled by the Hamiltonian
\begin{align}
H_\text{S}=\sum_{\nu=\lcirclearrowright, \rcirclearrowright} \varepsilon_0 d^\dagger_{\nu}d^{\phantom{\dagger}}_{\nu}+U d_{\lcirclearrowright}^{\dagger} d^{\phantom{\dagger}}_{\lcirclearrowright}d^\dagger_{\rcirclearrowright}d^{\phantom{\dagger}}_{\rcirclearrowright}\,,
\end{align}
where $d^\dagger_{\nu}$ and $d_{\nu}$ are creation and annihilation operators. The index $\nu\in\{\lcirclearrowright,\rcirclearrowright\}$ describes the valley degree of freedom, which for nanotubes with a zigzag edge corresponds to eigenstates with a clockwise ($\lcirclearrowright$) and counterclockwise  ($\rcirclearrowright$) rotation around the nanotube~\cite{jarillo_2005}. The real spin is not explicitly considered here and can be included by a simple degeneracy factor. The single-particle energy is denoted by $\varepsilon_0$ and the strength of the Coulomb repulsion by $U$. The eigenstates and eigenenergies are given by $\ket{\chi}\in\{\ket{0},\ket{\lcirclearrowright},\ket{\rcirclearrowright},\ket{\text{d}}\}$ and $E_{\chi}\in\{0,\varepsilon_0,\varepsilon_0,2\varepsilon_0+U\}$, respectively, where double occupation $\ket{\text{d}}$ is assumed to be irrelevant due to a strong Coulomb repulsion $U$. In Fig.~\ref{fig:setup}(b), the relevant states $\{\ket{0},\ket{\lcirclearrowright},\ket{\rcirclearrowright}\}$ are illustrated as an effective $\Lambda$ system.

To describe the dynamics of the quantum dot, we employ a master equation in Lindblad form for the reduced density matrix $\rho(t)$
\begin{align}\label{eq:push_master}
	\dot \rho={\cal L}\rho= &-i \left[ H_\text{S} +\delta H ,\rho\right] \nonumber \\
	&+\sum_{j} \Gamma_{j}\left(L_{j}^{\phantom{\dagger}}\rho L_{j}^{\dagger}-\frac{1}{2}\{ L_{j}^{\dagger} L_{j}^{\phantom{,\dagger}},\rho \}\right),
\end{align}
where $\delta H$ describes renormalization of the Hamiltonian, and $j$ labels all relevant Lindblad operators describing the coupling to the environment. In addition, we set $\hbar = 1$ and make use of the commutator $[\cdot,\cdot]$ and anticommutator $\{ \cdot,\cdot \}$. 

We start by modeling the coupling that drives electron transport and describes coherent population trapping.
To this end, we assume a tunnel-coupling Hamiltonian of the form 
\begin{align}\label{eq:tunnelcoupling}
    H_\text{C}=\sum_k  \left(t_\text{L}d^\dagger_{\text{L},+} c_{\text{L},k} +t_\text{R} c^\dagger_{\text{R},k}d_{\text{R},+} + \text{H.c.}\right),
\end{align}
with tunneling amplitudes $t_\text{L}$ and $t_\text{R}$ to source and drain, respectively. 
Here, $c_{l,k}$ and $c^\dagger_{l,k}$ with $l=\text{L,R}$ are the electron annihilation and creation operators of the environment with corresponding Hamiltonian $H_\text{E}=\sum_{l=\text{L,R}}\sum_k \varepsilon_k c^\dagger_{l,k}c_{l,k}$.   
The excitations in the system are described by the operators
\begin{subequations}
\begin{align}
d_{\text{L},\pm}^\dagger&=\frac{1}{\sqrt{2}} \Big( e^{i \phi_\text{L}} d^\dagger_{\lcirclearrowright}\pm e^{-i \phi_\text{L}} d^\dagger_{\rcirclearrowright}\Big), \label{eq:exc_left} \\
d_{\text{R},\pm}&=\frac{1}{\sqrt{2}} \Big( e^{-i \phi_\text{R}} d_{\lcirclearrowright}\pm e^{i \phi_\text{R}} d_{\rcirclearrowright}\Big).\label{eq:exc_right}
\end{align}
\end{subequations}
Thus, the two valleys $\ket{\lcirclearrowright}$ and $\ket{\rcirclearrowright}$ are coherently excited with equal weights but relative phases described by $\phi_\text{L}$ and $\phi_\text{R}$.
According to Eq.~\eqref{eq:tunnelcoupling}, only the excitations $d_{\text{L},+}^\dagger$ and $d_{\text{R},+}^{\phantom{\dagger}}$ play a role. The orthogonal excitations $d_{\text{L},-}^\dagger$ and $d_{\text{R},-}^{\phantom{\dagger}}$ are assumed to be \textit{dark}, i.e., decoupled from source and drain, respectively.
In Ref.~\cite{donarini_2019}, these dark excitations described by $\phi_\text{L}$ and $\phi_\text{R}$ are explained via the relative rotation of source and drain with respect to the carbon nanotube.

Using the coherent Lindblad approximation~\cite{kleinherbers_2020,kirisanskas_2018,Davidovic2020completelypositive,nathan_2020}, the tunnel coupling in leading order is described by the following Lindblad operators
\begin{subequations}
\begin{align}
L_{\text{L},\pm}&=\sum_{\chi,\chi^\prime} \sqrt{f(E_\chi-E_{\chi^\prime}-\mu_\text{L})}\mel{\chi}{d^\dagger_{\text{L},\pm}}{\chi^\prime} \dyad{\chi}{\chi^\prime} \nonumber\\ 
&\approx \sum_{\nu=\lcirclearrowright,\rcirclearrowright}\mel{\nu}{d^\dagger_{\text{L},\pm}}{0} \dyad{\nu}{0},\label{eq:lindblad1}\\
L_{\text{R},\pm}&=\sum_{\chi,\chi^\prime} \sqrt{1-f(E_{\chi^\prime}-E_{\chi}-\mu_\text{R})}\mel{\chi}{d_{\text{R},\pm}}{\chi^\prime} \dyad{\chi}{\chi^\prime} \nonumber\\
&\approx d_{\text{R},\pm} \label{eq:lindblad2} ,
\end{align}
\end{subequations}
where $f(\omega)=\left[e^{\omega/(k_\text{B} T)}+1\right]^{-1}$ is the Fermi-Dirac distribution, $T$ is the temperature, and $\mu_\text{L}$ and $\mu_\text{R}$ are the electrochemical potentials of the leads. To ensure unidirectional electron transport from left to right, we choose a small temperature $T$ and a large bias voltage $eV=\mu_\text{L}-\mu_\text{R}>0$ such that  $f(\varepsilon_0-\mu_\text{L})\approx 1$ and $f(\varepsilon_0-\mu_\text{R})\approx f(\varepsilon_0+U-\mu_\text{L/R})\approx 0$, see Fig.~\ref{fig:energies}.
Accordingly, the Lindblad operators describing electron transport from right to left (originating from the conjugated excitations $d_{\text{L},\pm}$ and $d_{\text{R},\pm}^\dagger$) are irrelevant.
The respective tunneling rates are given by
\begin{subequations}
\begin{align}
    \Gamma_{\text{L},+} &= 4\Gamma_\text{L},\\
    \Gamma_{\text{L},-} &= 0,\\
    \Gamma_{\text{R},+} &= 2\Gamma_\text{R},\\
    \Gamma_{\text{R},-} &= 0,
\end{align}
\end{subequations}
where $\Gamma_l=\pi \vert t_l\vert^2 D_l(\varepsilon_\text{F})$ with $l=\text{L,R}$ is given by the density of states $D_l(\varepsilon_\text{F})$ of source and drain at the Fermi energy $\varepsilon_\text{F}$.
The relative factor of two between $\Gamma_{\text{L},+}$ and $\Gamma_{\text{R},+}$ is due to the spin degree of freedom: while electrons can carry either spin up or down when tunneling in, the spin of the electron that tunnels out is given.

Due to the very nature of the coupling, there is a state  $\ket{\text{A}}$ that fully couples to the right lead, $\mel{0}{L_{\text{R},+}}{\text{A}}=1$, and a state $\ket{\text{B}}$ that fully decouples from it in terms of the excitation described by Eq.~\eqref{eq:lindblad2}, $\mel{0}{L_{\text{R},+}}{\text{B}}=0$. They are defined by
\begin{subequations}
\begin{align}
\ket{\text{A}}= d^\dagger_{\text{R},+} \ket{0} = \frac{1}{\sqrt{2}} \Big( e^{i \phi_\text{R}} \ket{\lcirclearrowright}+e^{-i \phi_\text{R}} \ket{\rcirclearrowright}\Big),\\
\ket{\text{B}}=d^\dagger_{\text{R},-} \ket{0} =\frac{1}{\sqrt{2}} \Big( e^{i \phi_\text{R}} \ket{\lcirclearrowright}-e^{-i \phi_\text{R}} \ket{\rcirclearrowright}\Big).
\end{align}
\end{subequations}
In contrast, the left lead excites (for $\phi_\text{L}\neq \phi_\text{R}$) a linear combination of $\ket{\text{A}}$ and $\ket{\text{B}}$.
Thus, with time, the system can evolve into state $\ket{\text{B}}$, which traps an electron and completely blocks the transport.  If $\ket{\text{B}}$ cannot be left directly, it is called a dark state. This case is also known as coherent population trapping.

To resume electron transport, the electron must somehow escape the trap.
In the following, we discuss three different physical mechanisms for this. 

\subsection{Decoherence}
One way to escape state $\ket{\text{B}}$ is via decoherence. Decoherence drives the state towards a total statistical mixture of $\ket{\text{B}}$ and $\ket{\text{A}}$.
Since electrons in state $\ket{\text{A}}$ can leave the dot, electron transport is resumed. Phenomenologically, this can be described by the Lindblad operators 
\begin{align}
L_k=I_k=\frac{1}{2}\sum_{\nu \nu^\prime}\big(\tau_{k}\big)_{\nu\nu^\prime}d^\dagger_{\nu}d^{\phantom{\dagger}}_{\nu^\prime},\end{align}
with rates $\Gamma_\text{dec}$. Here, we defined a valley spin operator $\vb{I}$ with components $I_k$ using the Pauli matrices $\tau_k$ with $k\in\{x,y,z\}$ in the basis $\ket{\lcirclearrowright}$ and $\ket{\rcirclearrowright}$. The isotropic combination of all three Lindblad operators $L_x$, $L_y$, and $L_z$ ensures that no particular state is preferred. Thus, on the Bloch sphere spanned by $\ket{\lcirclearrowright}$ and $\ket{\rcirclearrowright}$, the Lindblad operators drive each pure quantum state on the surface straight towards the center which corresponds to the fully mixed state. 

\subsection{Weak coupling}
Another possibility is that state $\ket{\text{B}}$ is actually not a fully dark but a very weakly coupled state --- a so called \textit{blocking state}~\cite{harabula_2018}. Thus there is a small probability that an electron leaves the quantum dot via the excitation $d_{\text{R},-}$ which is orthogonal to $d_{\text{R},+}$. This leads to modified tunneling rates for the Lindblad operators
\begin{subequations}
\begin{align}
    \Gamma_{\text{L},+} &= 4\Gamma_\text{L},\\
    \Gamma_{\text{L},-} &= 0,\\
    \Gamma_{\text{R},+} &= (2-x)\Gamma_\text{R},\\
    \Gamma_{\text{R},-} &= x \Gamma_\text{R}.
\end{align}
\end{subequations}
For $x=0$, state $\ket{\text{B}}$ is fully dark, while for $0<x\ll1$, it is a very weakly coupled blocking state. Here, we describe the weak coupling only phenomenologically, without specifying its microscopic origin. 

\subsection{Virtual charge fluctuations}
Finally, the third possibility to escape the trapping in state $\ket{\text{B}}$ is via virtual charge fluctuations with source and drain that modify the coherent dynamics of the system. Similar fluctuations have been described in Refs. \cite{braun_2004,schultz_quantum_2009,donarini_allelectric_2009}. They give rise to a correction of the Hamiltonian $H_\text{S}$ of the form
\begin{align}
\delta H = 2\,\vb{\boldsymbol{\omega}}\cdot \vb{I},
\end{align}
which describes a precession of the valley spin operator $\vb{I}$ around the vector $\boldsymbol{\omega} =\sum_{l=\text{L,R}}\omega_l(\cos(2\Delta \phi), \sin(2\Delta \phi),0)$.
By performing a spinful calculation including all two-electron states, we obtain for the energies $\omega_l$ with $l=\text{L,R}$ 
\begin{align}
\omega_l= \frac{\Gamma_l}{4} \bigg[&2R(\varepsilon_0-\mu_l)+R\left(\varepsilon_0+U-\frac{J}{2}-\mu_l\right)\nonumber\\
&-3R\left(\varepsilon_0+U+\frac{J}{2}-\mu_l\right) \bigg], \label{eq:renorm_spin}
\end{align}
where $\pi R(E)=\text{Re}\left[\Psi\left(\frac{1}{2} + i\frac{E}{2\pi k_\text{B}T} \right)\right]-\ln\left( \frac{W_c}{{2\pi k_\text{B} T}}\right)$ is given by the digamma function $\Psi(x)$~\cite{abramowitz_1988} and $W_c\gg k_\text{B} T$ is used for regularization. Here, the first, second, and third term describe virtual charge fluctuations to the empty, singlet, and triplet states, respectively.
However, since the exchange interaction $J$ is small~\cite{donarini_2019} compared to the Coulomb repulsion, $J\ll U$, we can approximate the energies as 
\begin{align}
\omega_l\approx \frac{\Gamma_l}{2}\bigg[&R(\varepsilon_0-\mu_l)-R\left(\varepsilon_0+U-\mu_l\right)\bigg],\label{eq:renorm_eff}
\end{align}
which exactly corresponds to the results of a spinless calculation. 

\section{kinetic equation}
\label{sec:kinetic}
Putting everything together, we can write the master equation
\begin{align}
	\dot{\rho}\z&={\cal L}\z\rho\z
\end{align}
as a matrix equation by combining the non-vanishing matrix elements of the density matrix into the vector $\rho=(\rho_0, \rho_\text{A}, \rho_\text{B},\rho^\text{A}_\text{B},\rho^\text{B}_\text{A})$. 
The superoperator ${\cal L}\z$ is, then, given by
\begin{widetext}
\begin{align}
    \label{eq:interactions_ds_wz}
	{\cal L}\z =
	\left(\begin{array}{ccccc}
	{-}4\Gamma_{\text{L}} & z(2-x)\Gamma_\text{R} & z\, x\, \Gamma_\text{R} & 0 & 0  \\  
4\Gamma_{\text{L}}\cos^2(\Delta \phi) &{-}(2-x)\Gamma_\text{R} {-}{\Gamma_{\text{dec}}}/{2}  & {\Gamma_{\text{dec}}}/{2} & -\omega_\text{L} \sin(2\Delta \phi) &-\omega_\text{L}\sin(2\Delta \phi)\\ 
	4\Gamma_{\text{L}}\sin^2(\Delta \phi)&{\Gamma_{\text{dec}}}/{2} &-x\,\Gamma_\text{R}{-}{\Gamma_{\text{dec}}}/{2} &\omega_\text{L}\sin(2\Delta \phi)&\omega_\text{L}\sin(2\Delta \phi) \\  
 -2i\Gamma_\text{L} \sin(2\Delta \phi) & \omega_\text{L} \sin(2\Delta \phi) & -\omega_\text{L} \sin(2\Delta \phi) & -\Gamma_\text{R}{-}\Gamma_\text{dec} {-}2i\tilde{\omega} & 0 \\
  2i\Gamma_\text{L} \sin(2\Delta \phi) & \omega_\text{L} \sin(2\Delta \phi) & -\omega_\text{L} \sin(2\Delta \phi) & 0 & -\Gamma_\text{R}{-}\Gamma_\text{dec} {+}2i\tilde{\omega}  \\
\end{array}\right),
\end{align}
\end{widetext}
where we used the abbreviations $\Delta \phi=\phi_\text{L}{-}\phi_\text{R}$ and $\tilde{\omega}=\omega_\text{R} +\omega_\text{L} \cos(2\Delta \phi)$.
We introduced the counting variable $z$ in the second and third column of the first row.
It keeps track of the number of electrons leaving the quantum dot \cite{kambly_2011,stegmann2015detection,kleinherbers2018revealing,stegmann_2016_short,ho_2019}.

In Eq.~(\ref{eq:interactions_ds_wz}), the phases $\phi_l$ only enter as the difference $\Delta \phi=\phi_\text{L}{-}\phi_\text{R}$ between the right and the left lead since we performed a phase transformation, $d_\nu \rightarrow U_\phi^\dagger d_\nu U_\phi$, with $U_\phi=\text{exp}\left[i\phi_\text{R}\left(d^\dagger_{\lcirclearrowright} d^{\phantom{\dagger}}_{\lcirclearrowright}-d^\dagger_{\rcirclearrowright} d^{\phantom{\dagger}}_{\rcirclearrowright}\right)\right]$.
The upper left $3\times 3$ matrix describes transitions between the populations of the states $\ket{0}$, $\ket{\text{A}}$, and $\ket{\text{B}}$, while coherences between $\ket{\text{A}}$ and $\ket{\text{B}}$ are covered by the fourth and fifth row and column. 
In the absence of the renormalization terms $\omega_l$, coherences between states $\ket{\text{A}}$ and $\ket{\text{B}}$ turn out to be irrelevant~\cite{donarini_2019,ho_2019}. 
Although such a coherence can be generated for $\sin(2\Delta \phi)\neq 0$ when an electron tunnels in (fourth and fifth row of the first column), for $\omega_\text{L}= 0$ this coherence only decays and does not affect the occupation probabilities of $\ket{0}$, $\ket{\text{A}}$, and $\ket{\text{B}}$ (first three rows of the fourth and fifth column).
We remark that the absence of $\omega_\text{R}$ in the off-diagonal matrix elements of ${\cal L}\z$ is due to the chosen basis states $\ket{\text{A}}$ and $\ket{\text{B}}$ adjusted to the right lead.

In the remaining part of the paper, we compare two different limits of Eq.~\eqref{eq:interactions_ds_wz}. First, we study a perfect dark state with $x=0$, where the electrons can leave the quantum dot only indirectly either by decoherence ($\Gamma_\text{dec}$) or coherently via virtual charge fluctuations ($\omega_\text{L}$).
Second, we study a blocking state with $x>0$ in the absence of decoherence, $\Gamma_\text{dec}=0$, such that the electron can leave the quantum dot either directly with rate $x\,\Gamma_\text{R}$ or indirectly via virtual charge fluctuations ($\omega_\text{L}$). Our goal is to find out whether it is possible to distinguish these two models within a transport measurement.

\subsection{Dark state}
\begin{figure}[h]
  \begin{center}
    \includegraphics[width=.3\textwidth]{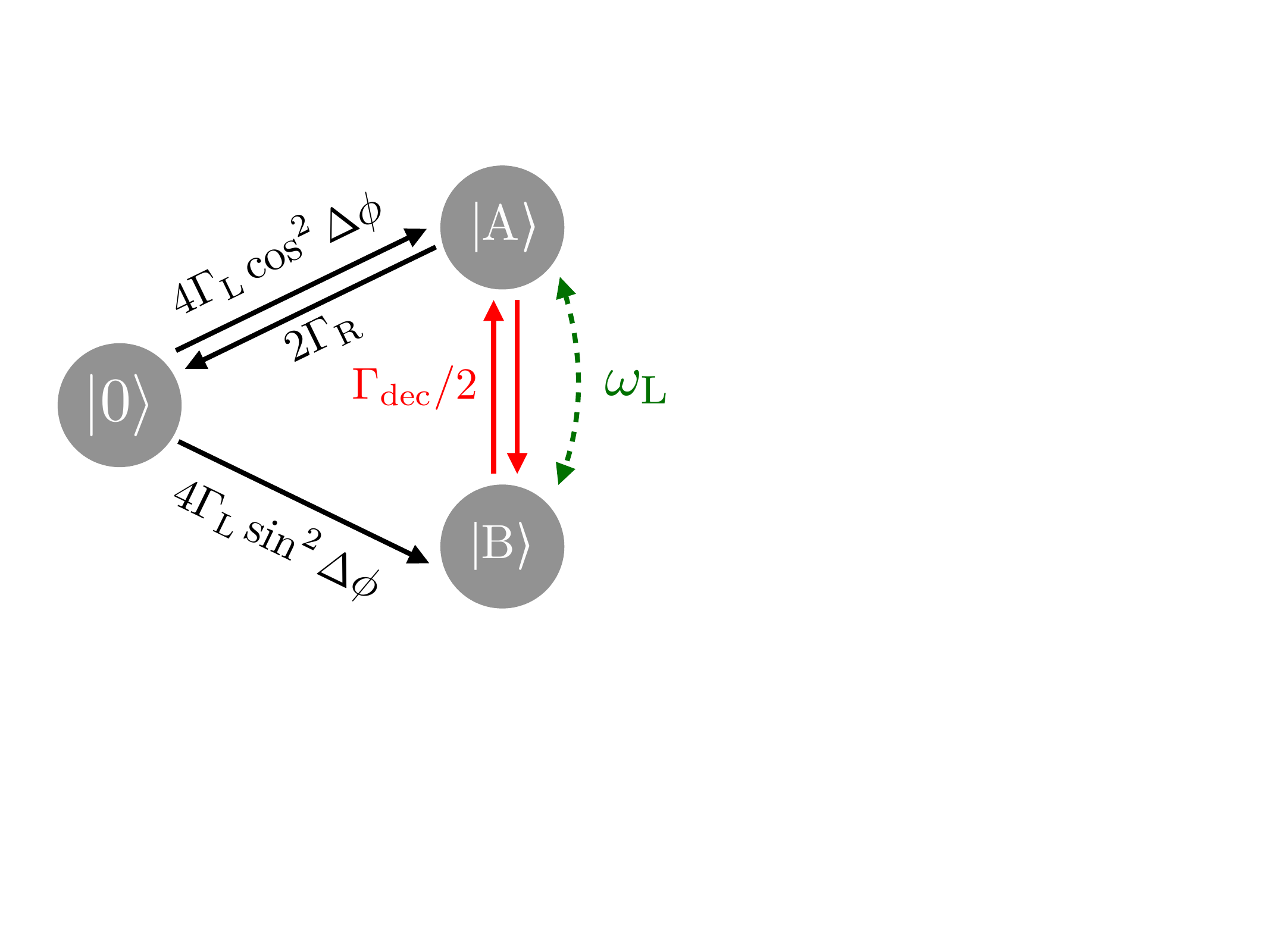}
  \end{center}
  \caption{Possible transitions in the dark-state model described by Eq.~\eqref{eq:interactions_ds_wz} with $x=0$ and $\Gamma_\text{dec}\neq 0$. Transitions where electrons enter or leave the quantum dot are indicated by black arrows.  Electrons can only leave the dark state $\ket{\text{B}}$ via the coupled state $\ket{\text{A}}$ (red arrows), and thus transport is suppressed. There are also transitions between $\ket{\text{A}}$ and $\ket{\text{B}}$ that are caused by virtual charge fluctuations with the left lead (dashed green arrow). 
  }
  \label{fig:sketchdark}
\end{figure}
For a perfect dark state, $x=0$, all possible transitions are indicated in Fig.~\ref{fig:sketchdark}. Upon neglecting renormalization due to virtual fluctuations, $\omega_\text{L}=\omega_\text{R}=0$, the current can be expressed simply as~\cite{ho_2019}
\begin{align}
\ev{I}=\left.\tr\left(\partial_z{\cal L}\z \rho_\text{st}\right)\right\vert_{z=1} =\frac{1}{\frac{2 \sin ^2(\Delta \phi )}{\Gamma_{\text{dec}}}+\frac{4 \Gamma_{\text{L}}+\Gamma_{\text{R}}}{4
\Gamma_{\text{L}} \Gamma_{\text{R}}}},
\end{align}
where $\rho_\text{st}$ is the vector of the density matrix elements in the steady state, found from ${\cal L}_1 \rho_\text{st}=0$. In the following, however, we include the renormalization effect, such that the formula for the current becomes more complicated.

In Fig.~\ref{fig:curr}, we show the current as a function of the phase difference $\Delta \phi$ of the couplings to the left and right lead. 
\begin{figure}[th]
  \includegraphics[width=.5\textwidth]{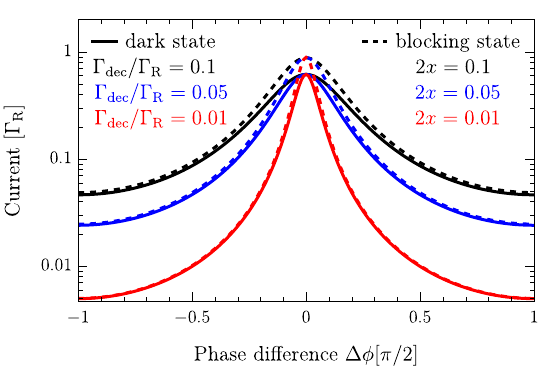}
  \caption{Electron current $\ev{I}$ as a function of the phase difference $\Delta \phi$. Here we compare the dark-state model, $\Gamma_\text{dec}>0$ and $x=0$ (solid lines), with the blocking-state model, $\Gamma_\text{dec}=0$ and  $x>0$ (dashed lines), for relaxation rates $\Gamma_\text{dec}=0.1\,\Gamma_\text{R}$ (black), $0.05\,\Gamma_\text{R}$ (blue), $0.01\,\Gamma_\text{R}$ (red) and tunnel couplings $2x=0.1$ (black), $0.05$ (blue), $0.01$ (red), respectively.
  We choose $\Gamma_\text{L}=0.4\,\Gamma_\text{R}$. To obtain $\omega_\text{L/R}$, we used $k_\text{B} T=50\,\mu \text{eV}$, $\varepsilon_0=0.75\,\text{meV}$, $U=20\,\text{meV}$, $\mu_\text{L}=\eta eV$, and $\mu_\text{R}=(\eta-1) eV$ with $\eta=0.55$ and $eV=19\,\text{meV}$ to comply with the situation in Ref.~\cite{donarini_2019}. This choice of parameters leads to $\omega_\text{L}=0.038\,\Gamma_\text{R}$ and $\omega_\text{R}=0.183\,\Gamma_\text{R}$.}
  \label{fig:curr}
\end{figure}
For zero phase difference, $\Delta \phi=0$, the current acquires its maximum value since the left and right lead fully couple to the same state,  $\mel{\text{A}}{L_{\text{L},+}}{0}=1$ and $\mel{0}{L_{\text{R},+}}{\text{A}}=1$. However, with an increasing mismatch, $\Delta \phi \neq0$, also a dark state $\ket{\text{B}}$ can be excited on the quantum dot and the electron transport becomes suppressed. This effect is most significant for phase differences of $\Delta \phi =\pi/2$, where  the left lead no longer excites the coupled state but only the dark state, $\mel{\text{A}}{L_{\text{L},+}}{0}=0$ and $\mel{\text{B}}{L_{\text{L},+}}{0}=1$.
We also find that as the decoherence rate $\Gamma_\text{dec}$ increases (from red to blue to black), the current increases as well. 
For $\Gamma_\text{dec}\rightarrow 0$, the current is maximally suppressed, with virtual charge fluctuation being the only source of a finite current. The parameters used for Fig.~\ref{fig:curr} are taken from the experiment in Ref.~\cite{donarini_2019}. The bias voltage $eV=\mu_\text{L}-\mu_{\text{R}}$ between source and drain is chosen in such a way that virtual charge fluctuations with the left lead are partially suppressed and decoherence is the dominant escape mechanism out of state $\ket{\text{B}}$.

\subsection{Blocking state}
\begin{figure}[h]
  \begin{center}
    \includegraphics[width=.3\textwidth]{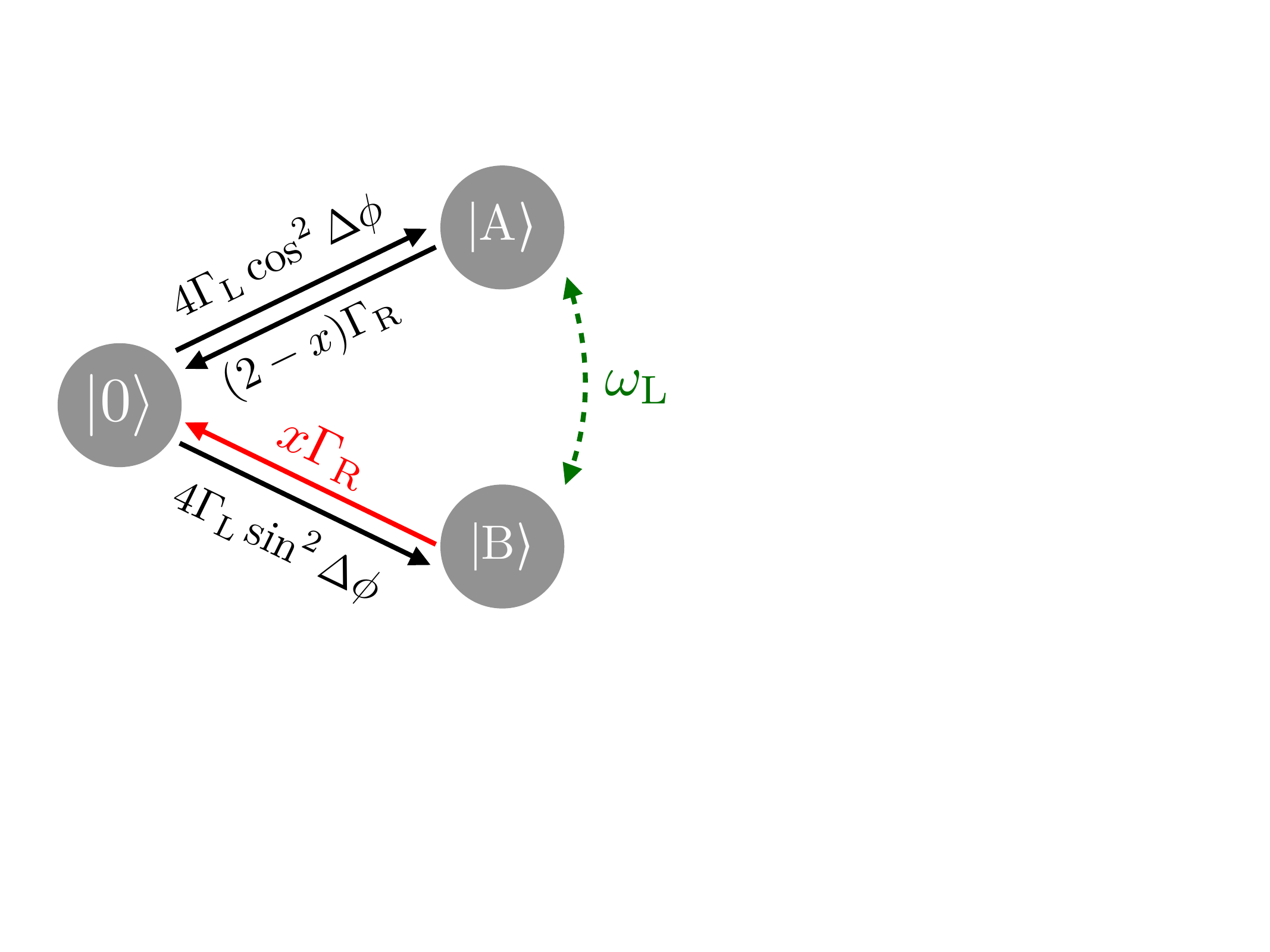}
  \end{center}
  \caption{Possible transitions in the blocking-state model described by Eq.~\eqref{eq:interactions_ds_wz} with $x>0$ and $\Gamma_\text{dec}= 0$. Electrons can enter the quantum dot in the same way as in the dark-state model (see black arrows), but the tunneling rates for electrons leaving the dot change. Electrons can now directly escape the blocking state $\ket{\text{B}}$ with rate $x \Gamma_\text{R}$ (red arrow). Therefore, the transport is suppressed for small $x$. Switching between the states $\ket{\text{A}}$ and $\ket{\text{B}}$ via decoherence is not possible anymore. The effect of the virtual charge fluctuations, on the other hand, is the same for the dark-state and the blocking-state model.}
  \label{fig:sketchblocking}
\end{figure}
The suppression of the electron transport due to a dark state is very similar to a situation where a so-called blocking state with $x>0$ is realized~\cite{harabula_2018}. 
A blocking state is a quantum state that couples only very weakly to the drain electrode with rate $x\Gamma_\text{R}$ with $x\ll1$.
This situation is sketched in Fig.~\ref{fig:sketchblocking}.
Similar as for the dark-state model, electron current is suppressed, see Fig.~\ref{fig:curr}. 
In this case, it is, however, not the decoherence but rather the direct coupling to the drain which allows for escaping the trapped state.

To allow for a fair comparison of the dark-state model ($x=0$ and $\Gamma_\text{dec}> 0$) with the blocking-state model ($x> 0$ and $\Gamma_\text{dec}= 0$), we choose $x$ and $\Gamma_\text{dec}= 0$ such that the respective rates for getting out of state $\ket{\text{B}}$ are the same, i.e.,
\begin{align}
\label{eq:x_dec}
2x\Gamma_\text{R}=\Gamma_\text{dec}.
\end{align}
With this choice, the values for the electric current for the two models become almost identical for phase differences of $\Delta \phi = \pi/2$, as can be seen in  Fig.~\ref{fig:curr}. 

\section{Factorial cumulants}
\label{sec:factorial_cumulants}

The comparison of Fig.~\ref{fig:sketchdark} with Fig.~\ref{fig:sketchblocking} shows that dark-state and blocking-state model differ qualitatively from each other.
On the other hand, this difference is not reflected in the behavior of the current, see Fig.~\ref{fig:curr}.
This raises the question whether it is possible to distinguish the two scenarios with a transport measurement at all and, if so, how this can be achieved.

Obviously, measuring the average charge current is not sufficient.
To circumvent this problem, we suggest to study electron transport in a time-resolved manner by monitoring the tunneling out of each individual electron as a function of time.
Recording such a time trace of tunneling events provides the maximal accessible information about the system's charge dynamics.
Statistical properties of the electron transport are fully included in full counting statistics, described by the probabilities $P_N(t)$ that $N$ electrons have left the quantum dot in a time interval $t$.
To analyze the distribution $P_N(t)$, we employ factorial cumulants~\cite{kambly_2011,stegmann2015detection,kleinherbers2018revealing} which can be obtained from the cumulant generating function 
\begin{align}
\label{eq:cum_gen}
{\cal S}(z,t)=\ln \tr\left( e^{{\cal L}\z t} \rho_\text{st}\right)
\end{align}
via derivatives $C_{\text{F},m}(t)=\partial_z^m {\cal S}(z)\vert_{z=1}$ with respect to the counting variable $z$.

A quantitative comparison of the full time dependence of measured factorial cumulants $C_{\text{F},m}(t)$ of orders $m=1,2,3\ldots$ with different theoretical models could, in principle, be used to rule out any proposed model.
There is, however, a qualitative and, hence, more stringent possibility that relies on the \textit{sign} of the factorial cumulant.
If electron transport is supported by uncorrelated tunneling events then the sign of each factorial cumulant is fixed by~\cite{kambly_2011,stegmann2015detection}
\begin{align}\label{eq:interactions_sign}
(-1)^{m-1}C_{\text{F},m}(t)\ge 0.
\end{align}
This, in turn, means that whenever this inequality is violated for any order $m$ at any time $t$, correlations must be present in the electron transfer.
Therefore, the sign of the factorial cumulants may serve as a suitable indicator of correlations and, in addition, may help to distinguish between different models, namely in cases in which different models predict different signs.

\begin{figure}[t]
  \includegraphics[width=.5\textwidth]{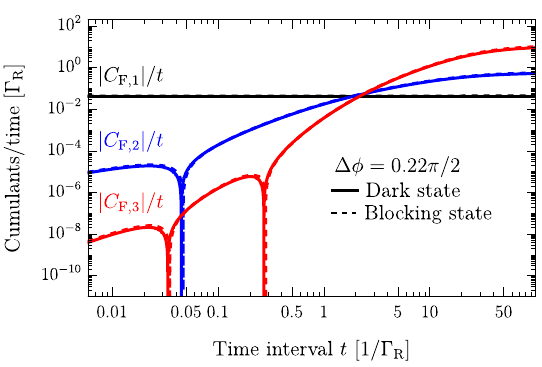}
  \caption{Factorial cumulants $C_{\text{F},m}$ as a function of time $t$ for the dark-state (solid) and blocking-state (dashed) model. The parameters are $\Gamma_\text{dec}=0.01\,\Gamma_\text{R}$ and $\Delta \phi=0.22\,\pi/2$. 
The remaining parameters are the same as in Fig.~\ref{fig:curr}.
}
  \label{fig:facs}
\end{figure}

In Fig.~\ref{fig:facs}, we display the time dependence of the first-, second- and third-order factorial cumulants for the dark-state (solid) and the blocking-state (dashed) model by using realistic parameters from the experiment~\cite{donarini_2019}. 
It is convenient to use a logarithmic scale, first, to present factorial cumulants of different order (whose values may differ by orders of magnitude) in one plot and, second, to display the power-law behavior $C_{\text{F},m}(t) \sim (-1)^{m-1}t^m$ at low $t$~\cite{stegmann_2016_short}.
In addition, we divide by $t$, so that at large $t$ the curves approach a constant.
Since negative values cannot be represented in a logarithmic plot, we show the modulus of the factorial cumulants only.
Nevertheless, we can identify sign \textit{changes} of the factorial cumulants as a function of time by the sharp spikes (in Fig.~\ref{fig:facs}, we find one sign change for the second and two sign changes for the third factorial cumulant). 

The sign changes visible in Fig.~\ref{fig:facs} clearly indicate that the electron transport is highly correlated. 
In fact, already the second factorial cumulant violates the inequality $-C_{\text{F},2}(t)\le 0$, which corresponds to a super-Poissonian Fano factor, associated with the fact that electrons are effectively transferred in bunches.
While this is interesting as such, we find that both the dark-state and the blocking-state model are basically indistinguishable even for higher-order factorial cumulants.
From this we conclude that even with full counting statistics, a distinction between the two models is not possible for the parameters given in the experiment~\cite{donarini_2019}.

Is this a generic statement or does it depend on the system parameters?
Or, to put it differently, can one suggest a change of system parameters such that the dark-state and the blocking-state model predict different signs of the factorial cumulants?
Instead of unsystematically calculating many factorial cumulants for a large parameter space, we aim at a more systematic approach to identify regions in which the two models can be distinguished by full counting statistics.
This procedure is based on the analytic structure of the cumulant generating function Eq.~(\ref{eq:cum_gen}) in the complex plane, as explained in the next section.

\section{Lee-Yang zeros}\label{sec:lee_yang}

\begin{figure}[t]
  \includegraphics[width=.48\textwidth]{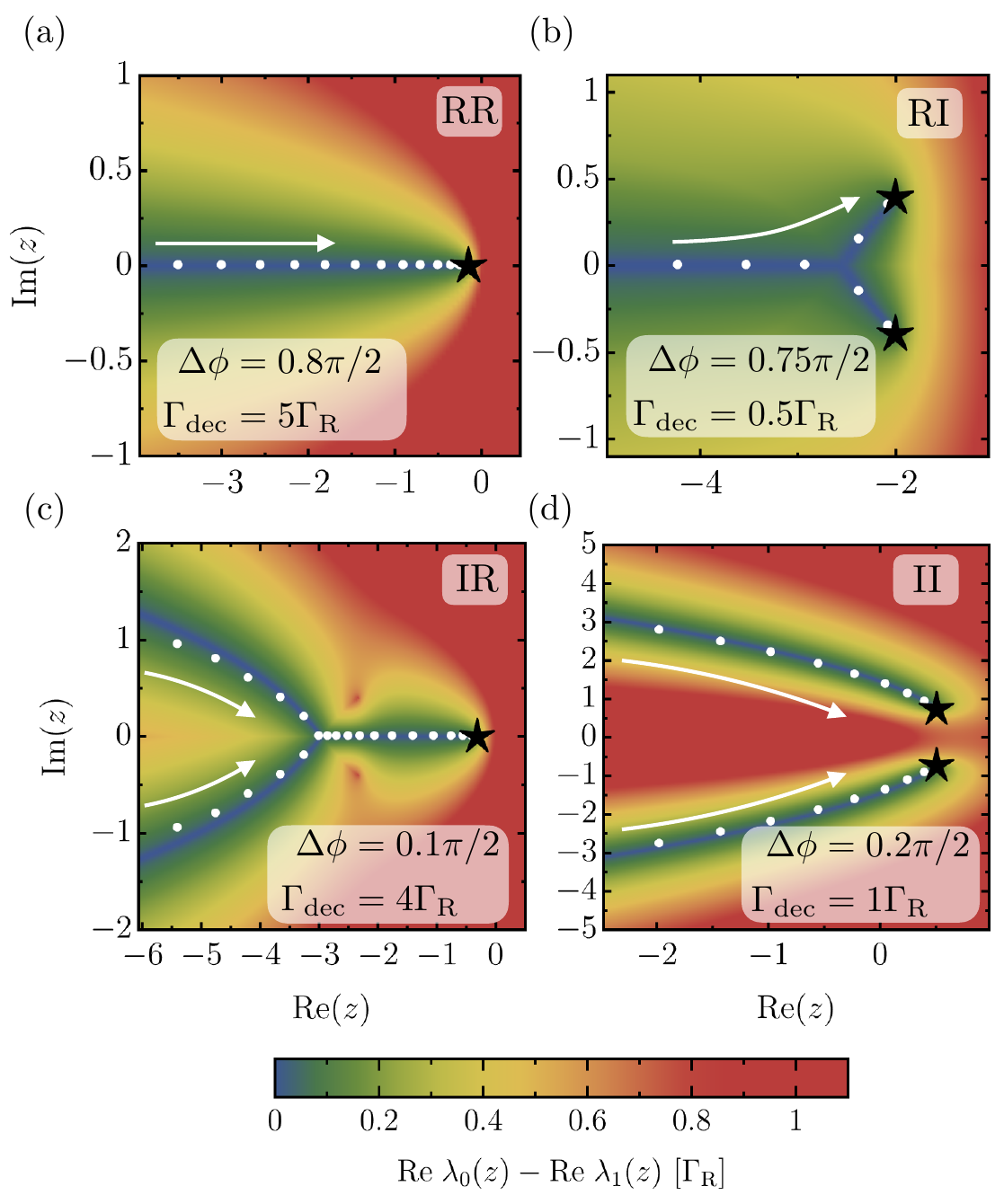}
  \caption{Lee-Yang zeros (white dots) in the complex plane for various combinations of parameters $\Delta \phi$ and $\Gamma_\text{dec}$ obtained within the dark-state model at $\Gamma_\text{R}t=20$. The difference $\text{Re}\, \lambda_0(z)-\text{Re}\, \lambda_1(z)$ is represented by the background color with blue background indicating a vanishing difference. (a)~For  $\Delta \phi=0.8\, \pi/2$ and $\Gamma_{\text{dec}}=5\,\Gamma_{\text{R}}$, all zeros are on the real axis (RR). (b)~For $\Delta \phi=0.75\,\pi/2$ and $\Gamma_{\text{dec}}=0.5\,\Gamma_{\text{R}}$, only the zeros close to $z_\text{A}$ leave the real axis (RI). (c)~For $\Delta \phi=0.1\,\pi/2$ and $\Gamma_{\text{dec}}=4\,\Gamma_{\text{R}}$, only the zeros far away from the origin,  $z\rightarrow z_{-}$, leave the real axis (IR). (d)~For $\Delta \phi=0.2\,\pi/2$ and $\Gamma_{\text{dec}}=1\,\Gamma_{\text{R}}$, all zeros leave the real axis (II). See Sec.~\ref{sec:arrangements in complex plane} for a more detailed explanation of this categorization. The remaining parameters are the same as in Fig.~\ref{fig:curr}.}
  \label{fig:zeros}
\end{figure}

Lee-Yang zeros are a concept from statistical physics and originally refer to the zeros of a grand-canonical partition function as a function of the fugacity. Lee-Yang theory was developed as a tool to reveal phase transitions in the Ising model \cite{yang_52,lee_52} and was later generalized to other systems \cite{asano_generelized_68,asano_leeyang_70,ruelle_extension_70,biskup_general_00,brandner_2017,deger_leeyang_20,brange2024lee}.

In the context of full counting statistics, the equivalent to the grand-canonical partition function is the moment generating function
\begin{align}
{\cal G}(z,t)= e^{{\cal S}(z,t)} = \sum_N z^N P_N(t) ,
\end{align}
which can be factorized~\footnote{In an actual experiment, there is always a maximum number of counts $N_\text{max}$ for any finite time $t$, so that ${\cal G}(z,t)$ is a finite polynomial of order $N_\text{max}$. Then, factorization is always ensured by the fundamental theorem of algebra.},
\begin{align}
{\cal G}(z,t)= \prod_n \frac{  z-z_n(t)}{1-z_n(t)},
\end{align}
where the denominator ensures normalization, ${{\cal G}(1,t)=1}$.
The zeros $z_n(t)$ of ${\cal G}(z,t)$ are referred to as dynamical Lee-Yang zeros~\cite{brandner_2017}.
They are either real or occur as complex conjugate pairs because of ${\cal G}^*(z,t)={\cal G}(z^*,t)$.
For the cumulant generating function ${\cal{S}}(z,t)$, the $z_n(t)$ are poles.
We obtain
\begin{align}\label{eq:interactions_pseudopoissbin}
{\cal{S}}(z,t)=\sum_n\ln\Big\{ \left[1-\tilde{p}_n(t)\right] +z\,\tilde{p}_n(t)\Big\},
\end{align}
with $\tilde{p}_n=1/(1{-}z_n)$.
The $\tilde{p}_n$ can be interpreted as \textit{pseudo-probabilities}. 
If they are real and lie between $0$ and $1$ then the cumulant generating function is that of a Poisson binomial process with $\tilde{p}_n$ being the probability that the transition $n$ takes place, which is independent of all other transitions~\cite{stegmann_2016_short}.
If, on the other hand, the pseudo-probabilities $\tilde{p}_n$ (and, thus, the Lee-Yang zeros $z_n$) have a finite imaginary part, it is no longer possible to understand the stochastic process in terms of uncorrelated transitions.
Instead, correlations must be present~\cite{kambly_2011,stegmann2015detection} and there is a chance to reveal them by a violation of the inequality Eq.~\eqref{eq:interactions_sign}.

\begin{figure*}[t]
  \includegraphics[width=.9\textwidth]{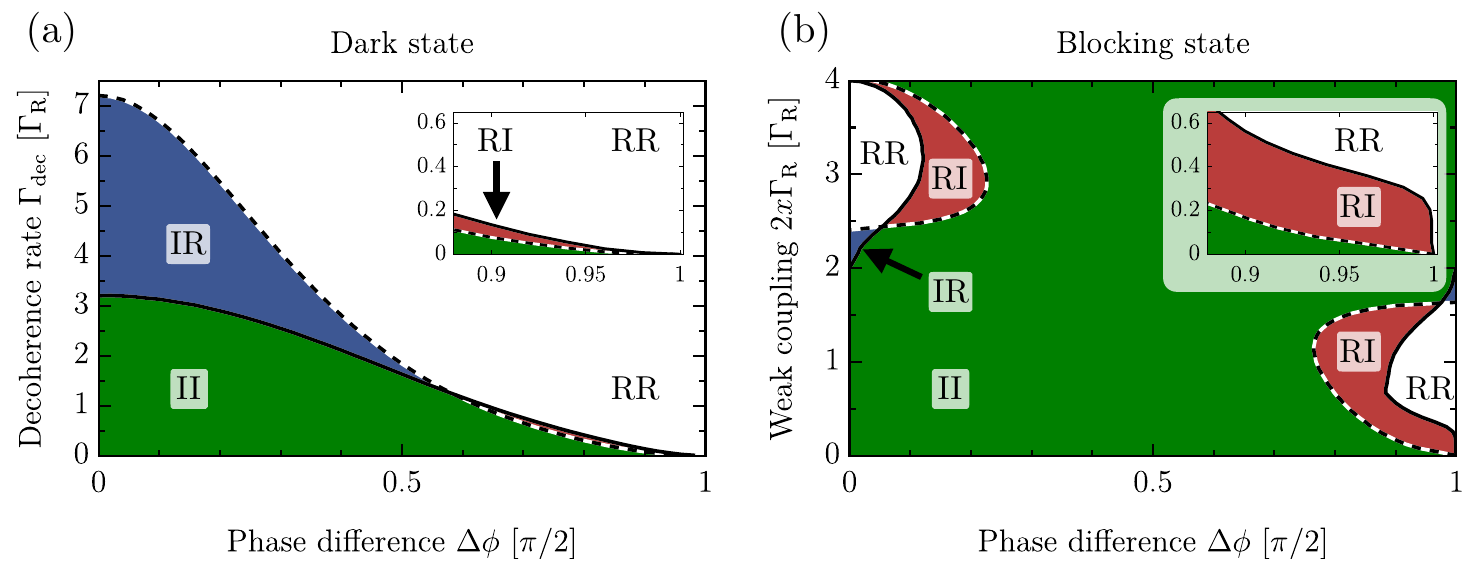}
  \caption{Parameter space of (a)~dark-state and (b)~blocking-state model. The parameters are phase difference $\Delta \phi$ and (a) the decoherence rate $\Gamma_\text{dec}$ for the dark-state model or (b) the coupling $x$ for the blocking-state model. For a comparison of (a) and (b), we choose $2x\Gamma_\text{R}=\Gamma_\text{dec}$. The regimes RR, RI, IR, and II identified in Sec.~\ref{sec:classification} and Fig. \ref{fig:zeros} are colored in white, red, blue, and green, respectively. The remaining parameters are the same as in Fig.~\ref{fig:curr}.
  }
  \label{fig:regimes}
\end{figure*}

\subsection{Classification of different correlation regimes}
\label{sec:classification}
In Fig.~\ref{fig:zeros}, we show the position of the Lee-Yang zeros as white dots in the complex plane for the dark-state model and some finite length $t$ of the time interval.
They are not randomly distributed but form some regular pattern that depends on the system parameters.
In panel (a), they all lie on the negative real axis.
In this case, the stochastic model can be interpreted in terms of a Poisson binomial distribution of uncorrelated tunneling events.
This contrasts with the scenarios depicted in panels (b)-(d), in which at least some of the Lee-Yang zeros have a finite imaginary part, which indicates correlations.
In panel (b), only the Lee-Yang zeros on the right, in (c) only those on left, and in (d) all of them have a finite imaginary part.
With increasing time $t$, every Lee-Yang zero moves from left to right (white arrows) to an attraction point $\bp$ (black stars in Fig.~\ref{fig:zeros}).
In the limit $t\to \infty$, they form a dense set on a contour line denoted as $\cal M$ which is indicated by blue color in Fig.~\ref{fig:zeros}.

To find the asymptotic Lee-Yang zero behavior, we perform, following Ref.~\cite{kambly_2011}, a spectral decomposition of the Liouvillian ${\cal L}\z$ and retain in the expression for the moment generating function only the contributions coming from the two eigenvalues $\lambda_0(z)$ and $\lambda_1(z)$ with the largest real parts, $\text{Re}\,\lambda_0\ge \text{Re}\,\lambda_1 > \text{Re}\,\lambda_{i\neq 0,1}$.
This yields
\begin{equation}
{\cal G}(z,t) \approx a_0(z) e^{\lambda_0(z) t} +a_1(z) e^{\lambda_1(z) t}.
\end{equation}
Contributions from eigenvalues with smaller real part can be neglected for large $t$.
The condition ${\cal G}(z,t)=0$ leads to 
\begin{equation}\label{eq:bp}
\lambda_0(z)-\lambda_1(z) =\frac{\ln(a_1/a_0)+i \pi (2n+1)}{t},
\end{equation} 
(see also Ref.~\cite{kambly_2013}), where $n$ is some integer. The real part of Eq.~(\ref{eq:bp}) in the limit $t\rightarrow \infty$ gives rise to
\begin{equation}\label{eq:contour}
\text{Re}\,\lambda_0(z)-\text{Re}\,\lambda_1(z) =0,
\end{equation} 
which defines the contour line $\cal M$. The imaginary part of Eq.~(\ref{eq:bp}), on the other hand, determines the individual positions of the Lee-Yang zeros on the contour $\cal M$, where each $n$ determines a given $z_n(t)$. To categorize the arrangements of Lee-Yang zeros seen in Fig. \ref{fig:zeros}, we study both the attraction point $\bp$ of the contour $\cal M$ and its behavior in the limit $\text{Re}\, z \rightarrow - \infty$.

\subsubsection{Attraction points of Lee-Yang zeros}

For $t\rightarrow \infty$, every zero $z_n(t)$ of the generating function approaches the solution of 
\begin{align}\label{eq:bpcond}
\lambda_0(\bp)=\lambda_1(\bp).
\end{align}
This equation uniquely determines the attraction points $\bp=\bp\,(\Gamma_\text{dec},\Delta \phi)$ as a function of the system parameters. They are typically square-root branch points and occur as complex conjugate pairs [see stars in Figs.~\ref{fig:zeros}(b) and \ref{fig:zeros}(d)] or are completely real [see stars in Figs.~\ref{fig:zeros}(a) and \ref{fig:zeros}(c)]. 
In cases where Eq.~(\ref{eq:bpcond}) cannot be solved analytically, we use the approximate method to find the position of $z_\text{A}$ presented in Ref.~\cite{kambly_2011} (based on Refs.~\cite{Zamastil_2005,flindt_2010}) that makes use of factorial cumulants of high order, $m\gg 1$.

\subsubsection{Lee-Yang zeros at $\text{{\normalfont Re}}\, z \rightarrow -\infty$}
To determine whether the Lee-Yang zeros $z_-$ in the limit $\text{Re}\, z \rightarrow -\infty$ are real or have a nonzero imaginary part, we use Eq.~\eqref{eq:contour} far from the origin. The eigenvalues $\lambda_i(z)$ in the limit $z\to -\infty$ can be calculated analytically by entering the \textit{ansatz} of a Puiseux series
\begin{align}
    \label{eq:series}
    \lambda_i(z)=b_{1,i}\sqrt{z}+b_{0,i}+b_{-1,i}\frac{1}{\sqrt{z}}+\ldots
\end{align}
into the characteristic polynomial of ${\cal L}_z$. Inserting this into Eq. (\ref{eq:contour}) at $ \text{Re}\, z \rightarrow -\infty$ and $ \text{Im}\,z = 0$  leads to the conditions 
\begin{subequations}
\begin{align}
    \text{Im}\, b_{1,1} &= \text{Im}\, b_{1,0},\\
    \text{Re}\, b_{0,1} &= \text{Re}\, b_{0,0},
\end{align}
\end{subequations}
for the coefficients of Eq. (\ref{eq:series}).
The zeros $z_-$ only lie on the real axis if both conditions are met. This allows us to determine for which system parameters the imaginary part of the Lee-Yang zeros $\text{Im}\, z_{-}(\Gamma_\text{dec},\Delta \phi)$ are finite. For $\text{Im}\, z_{-}=0$, the zeros far away from the origin are real [see Fig.~\ref{fig:zeros}(a)-(b)] while for $\text{Im}\, z_{-}\neq 0$ they occur as complex conjugate pairs [see Fig.~\ref{fig:zeros}(c)-(d)].

\subsubsection{Arrangements of Lee-Yang zeros in the complex plane}
\label{sec:arrangements in complex plane}
The four examples shown in Fig.~\ref{fig:zeros} correspond to four topologically different arrangements of the Lee-Yang zeros labeled by $\alpha\beta$, with $\alpha,\beta \in \left\{\text{R,I}\right\}$. 
Here, the first label $\alpha$ indicates whether the Lee-Yang zeros on the left $z_-$ are real (R) or have a finite imaginary part (I).
Similarly, the second label $\beta$ indicates the absence (R) or presence (I) of a finite imaginary part of the Lee-Yang zeros on the right, i.e., of $z_\text{A}$. Figures \ref{fig:zeros}(a), \ref{fig:zeros}(b), \ref{fig:zeros}(c), and \ref{fig:zeros}(d) correspond to configurations RR, RI, IR, and II, respectively.

\begin{figure*}[t]
  \includegraphics[width=.9\textwidth]{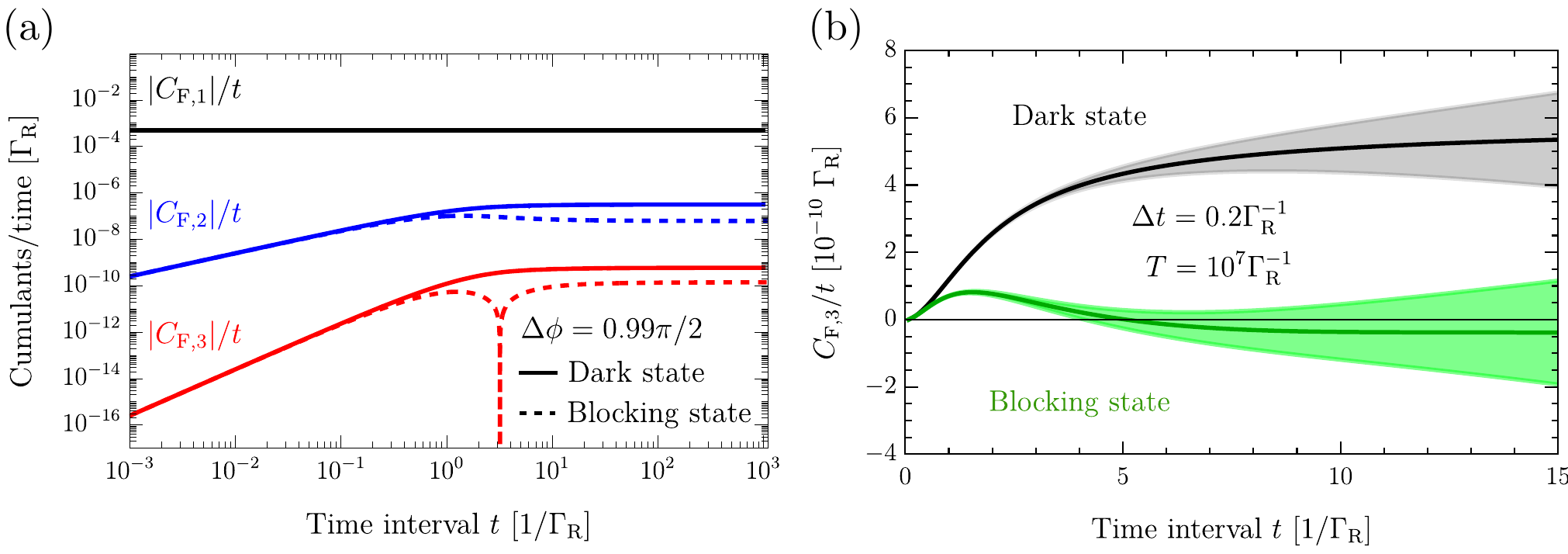}
  \caption{(a) Factorial cumulants $C_{\text{F},m}$ as a function of time $t$ for the dark-state (solid) and blocking-state (dashed) model.   (b) Third factorial cumulant $C_{\text{F},3}$ for the dark-state (black) and the blocking-state (green) model. Statistical errors (shaded background) due to a finite time trace ($T=10^7\,\Gamma_\text{R}^{-1}$) and systematic errors due to a limited time resolution ($\Delta t=0.2\,\Gamma_\text{R}^{-1}$) are included. The parameters are $\Delta \phi=0.99\,\pi/2$ and $\Gamma_\text{dec}=0.001$. The remaining parameters are the same as in Fig.~\ref{fig:curr}.
}
  \label{fig:interactions_DSBS_faccum_res}
\end{figure*}

\subsection{Parameter spaces for dark-state and blocking-state model}

This classification can be used to systematically analyze the parameter space of the dark-state and blocking-state model with respect to correlations, see Fig.~\ref{fig:regimes}. We vary the phase difference $\Delta \phi$ and the rate with which the dark or blocking state are left.
In white we show the regime RR, for which all Lee-Yang zeros remain on the real axis.
The other regimes are colored, II in green, IR in blue, and RI in red. Transitions from $\alpha \text{R}$ to $\alpha \text{I}$ are indicated by solid lines and transitions from $\text{R}\beta$ to $\text{I}\beta$ by dashed lines.

For both the dark-state and the blocking-state model, the characteristic polynomial of the superoperator ${\cal L}_z$ is $\pi$-periodic and even in $\Delta \phi$.
Therefore, in Fig.~\ref{fig:regimes}, we only show  the parameter regimes $0\le \Delta \phi \le \pi/2$.
In the dark-state model, the decoherence rate can be assumed to be arbitrarily large. 
This contrasts with the blocking-state model, for which values of $x$ larger than $2$ are unphysical, as they would imply a negative value of the rate $\Gamma_{\text{R},+}$.
We, therefore, show the parameter space up to $2x\Gamma_\text{dec}=4\Gamma_\text{R}$ only. 
(In fact, already for half that value, $2x\Gamma_\text{dec}=2\Gamma_\text{R}$, both states $\ket{\text{A}}$ and $\ket{\text{B}}$ are equally strongly coupled to the drain, and one can hardly call this a blocking-state model anymore.)

For infinitesimally small decoherence, the dark-state model is always in the II regime (except for the singular point $\Delta \phi=\pi/2$).
With increasing decoherence rate, more and more Lee-Yang zeros move to the negative real axis until, for sufficiently large $\Gamma_\text{dec}$ (that depends on the phase difference $\Delta \phi$), the regime RR is realized. 

Also for the blocking-state model, the regime II dominates in a large area of the parameter space. 
In particular for infinitesimally small couplings of the blocking state the regime II is always realized.
This implies that for the parameters realized in the experiment of Ref.~\cite{donarini_2019} (small decoherence rate and moderate phase difference), both the dark-state and the blocking-state model are in the same correlation regime such that it is not possible to distinguish them via the sign of factorial cumulants.
However, by inspecting the lower right corner of the parameter space in Fig.~\ref{fig:regimes} (see also the inset), we find that for a phase difference $\Delta \phi$ close to $\pi/2$, a distinction between the two models by means of factorial cumulants seems to be possible.

\subsection{Distinguishing a dark from a blocking state with factorial cumulants}

While for a small phase difference $\Delta \phi$, a large decoherence rate is required in the dark-state model to suppress correlations in transport, there is an interesting region around $\Delta \phi=\pi/2$, in which small decoherence rates are sufficient to change from the regime II via RI to RR, see inset in panel (a) of Fig.~\ref{fig:regimes}.
Comparing the insets of panels (a) and (b), we find that there are parameters for which the dark-state model is in the RR regime while in the blocking-state model the RI regime is realized.

In Fig.~\ref{fig:interactions_DSBS_faccum_res}(a), we show the factorial cumulants for both the dark-state and the blocking-state model for parameters $\Gamma_\text{dec}=0.001\,\Gamma_\text{R}$ and $\Delta \phi=0.99 \, \pi/2$. 
In this case, the dark-state model is in the RR regime and, indeed, the factorial cumulants $C_{\text{F},m}$ (solid lines) always obey the inequality~(\ref{eq:interactions_sign}). 
As a consequence, no sign change as a function of time is visible.
In contrast, the blocking-state model (dashed lines) is in the RI regime, for which correlations occur and, indeed, we find that the third cumulant changes its sign as indicated by a sharp spike.

In the calculation of the factorial cumulants, we assumed a perfect detector with no false or missing events as well as an infinitely long time trace. 
In reality, however, some tunneling events are missed by the detector because of a finite time resolution $\Delta t$. 
This leads to systematic errors.
In addition, a finite measurement time $T$ leads to stochastic errors.
These effects can be simulated within an error model, as described in Ref.~\cite{kleinherbers_2021_pushing}.
The resulting third factorial cumulant $C_{\text{F},3}$ is shown for some typical values of $\Delta t$ and $T$ in panel (b) of Fig.~\ref{fig:interactions_DSBS_faccum_res}.
The solid lines include the systematic errors due to missed events, and the shaded area accounts for stochastic errors.
We conclude that despite the errors, a distinction of the dark-state from the blocking-state model is still possible with the help of the sign of factorial cumulants.

\section{Conclusions}\label{sec:conclusions}

Motivated by a recent experiment~\cite{donarini_2019}, we investigated the all-electric analog of coherent population trapping in carbon-nanotube quantum dots.
Whenever a so-called dark state is excited in the quantum dot, electron transport is suppressed since the trapped electron can only leave with the help of decoherence.
Suppression of transport could, however, also be due to a so-called blocking state, a state that is only weakly tunnel coupled to the drain electrode.
This raises the fundamental question of whether and how it is possible to distinguish the two scenarios within a transport measurement.

Going beyond measuring the average current through the carbon nanotube, we suggest to study the full counting statistics of charge transfer with the help of factorial cumulants, since they are sensitive to correlations.
In particular, a sign change of factorial cumulant as a function of time indicates the presence of correlations.

We use Lee-Yang zeros to distinguish between different correlation regimes.
In this way, we find that a clear distinction between the dark-state and the blocking-state model using the sign of the factorial cumulants is only possible for a large phase difference of the tunnel coupling to the source and drain electrode. 
For small values of the phase difference, both models have qualitatively the same full counting statistics.

Our Lee-Yang zero analysis makes a statement about possible parameters for which a dark state can be distinguished from a blocking state and for which the same qualitative behavior is expected.
It should be emphasized, however, that it neither rules out the existence of dark states in carbon-nanotube quantum dots nor does it provide any mechanism for a blocking state or assess how likely it is to exist. 
We merely understand our proposal as a challenge to experimentally realize a system in which one of the two competing models can be ruled out by analyzing the full counting statistics of electron transport in terms of factorial cumulants.

\begin{acknowledgments}
This work was supported by the Deutsche Forschungsgemeinschaft (DFG, German  Research Foundation) under Project-ID 278162697 -- SFB 1242, German National Academy of Sciences Leopoldina under Project No. LPDS 2019-10 and the Mercator Research Center Ruhr (MERCUR) under Project No. Ko-2022-0013. We thank A. Hucht for helpful discussions.
\end{acknowledgments}

\bibliography{bibliography}

\end{document}